\documentclass [12pt]{article}
\usepackage {graphicx}
\usepackage {longtable}
\def \C{\cal C}
\def \P{\cal P}
\def \T{\cal T}

\begin{document}
\setlength{\baselineskip}{10pt}
\title{\vskip -2.5 cm A selection rule for transitions in PT-symmetric quantum theory}
\author { {Lawrence R. Mead} \footnote {Communications to lawrence.mead@usm.edu}\\
{Dept. of Physics and Astronomy} \\{University of Southern Mississippi}\\
{Hattiesburg, MS 39406}\\ \\
{David Garfinkle} \\ {Dept. of Physics} \\
{Oakland University}\\{Rochester Hills, MI 48309}\\
{and Michigan Center for Theoretical Physics}\\
{Randall Laboratory of Physics}\\
{University of Michigan}\\
{Ann Arbor, MI 48109-1120}} 
\date{{\small \today}}
\maketitle
\begin{abstract}
Carl Bender and collaborators have developed a quantum theory
governed by Hamiltonians that are PT-symmetric rather than Hermitian.
To implement this theory, the inner product was redefined to
guarantee positive norms of eigenstates of the Hamiltonian. 
In the general case, which includes arbitrary 
time-dependence in the Hamiltonian, a modification of the Schr\"odinger equation
is necessary as shown by Gong and Wang to conserve probability. In this paper,
we derive the following selection rule: transitions induced by time dependence 
in a PT-symmetric Hamiltonian cannot occur between normalized states of 
differing PT-norm. We show three examples of this selection rule in action: two matrix 
models and one in the continuum.   
\end{abstract}
\medskip
\newpage
\setlength{\baselineskip}{16pt}
\section{Introduction}
\medskip
\hskip 25pt In 2002, Carl Bender, Dorje C. Brody and Hugh F. Jones discovered 
that the Hamiltonian in quantum theory need not be Hermitian~\cite{CB1}, 
provided that it be PT-symmetric. Here, P is parity and T is time reversal.
In such a theory, the inner product is redefined such as to produce an always 
positive norm for any eigenstate of the Hamiltonian. This type of quantum
theory is understood and has been studied extensively by Bender and 
collaborators as well as many other authors~\cite{A}.  
PT-symmetric quantum theory is now a well-established part of physics.
   
A standard and important calculation in quantum mechanics is to add a time 
dependent interaction to a time- independent Hamiltonian and calculate the 
probability of transitions between the energy states of the initial Hamiltonian whose eigenstates are presumed to be completely known. For example, consider a 
hydrogen atom which at some time encounters an electromagnetic wave. Should the atom be in its ground state, one may compute (in some order of perturbation 
theory) the probability that at some time in the future the atom will absorb 
energy from the field and make a transition to a higher energy state (resonant
absorption). If the atom is in an excited state, one is interested in 
calculating the probability that the atom is de-excited and thereby emits a 
corresponding electromagnetic wave (stimulated emission). In this case,
there results a selection rule that the angular momentum change of the
atom must be $\Delta \ell = \pm 1$ in order to globally conserve angular 
momentum. 
Additional important examples of transitions come to mind, such as a spin flip
in systems during magnetic or electron-spin resonance. 

In this paper, we will address the corresponding calculations of transition
probabilities in non-relativistic PT-symmetric quantum theory. We will
derive a {\it general} theorem, a selection rule, valid for a PT-symmetric
Hamiltonian with {\it arbitrary} time-dependence. The paper is organized
as follows. Section II presents basic properties of both time
independent and time dependent PT symmetric quantum theory and gives the 
notation to be used in the rest of the paper. In Section III we
derive our main result, and in Section
IV we give three examples: a matrix model having no transitions and one
which does, both illustrating the derived selection rule; then an example
of a continuum model for which exact calculations may be made.
We end with suggestions and questions for further work.
\section{PT-symmetric quantum theory}
The Hamiltonian must be PT-symmetric that is, 
\begin{equation}
[H,{\P \T} ]= 0,
\end{equation}
and in the {\it unbroken phase}: that
is, all of the energy eigenstates of $H$ must also be eigenstates of $\P \T$.  It then follows that
all the eigenvalues of $H$ are real. As we illustrate in a later section, the unbroken phase condition may 
restrict the allowed range of parameters defining the Hamiltonian. 
Here, the operator $\P$ is the standard parity operator: under parity,
$p \to -p$, $x \to -x$. $\T$ is time reversal: under time reversal,
$p \to -p$ and $i \to -i$. The latter requirement of complex conjugation
is necessary to preserve the fundamental position-momentum commutation
relation $[x,p]=i$ under the combined operation $\P\T$ (we set $\hbar=1$).

Evolution using the Schr\"odinger equation with a non-Hermitian Hamiltonian generally 
does not preserve the usual $L^2$ norm $<\psi|\psi>$.  Thus PT-symmetric quantum mechanics
requires the introduction of a new norm.  The PT-symmetric Schr\"odinger equation {\emph does} preserve
the PT norm $<\psi|\P|\psi>$.  However, the PT norm is not positive definite.  The solution is to 
introduce the operator $\C$ which is defined as follows: the eigenstates of $\C$ are the same as those of 
the Hamiltonian, but the eigenvalues are $\pm 1$ where the plus sign is used if the state has positive PT norm and the negative sign is used if the state has negative PT norm.  The (positive definite) norm used in PT-symmetric
quantum mechanics is then
\begin{equation}
\langle \psi | W | \psi \rangle = \langle \psi | \P\C | \psi \rangle.  
\end{equation}
It follows from the definition of the $\C$ operator that it has the following properties:
\begin{eqnarray}
[H,{\C}]=0 \cr
[{\P\T,C}]=0 \cr
{{\C}^2}=1
\end{eqnarray}
These properties are often used to produce a perturbative series expression for $\C$.~\cite{CB2,CB3}  

So far, we have spoken only about Hamiltonians which are independent of time.
In order to obtain information about transitions, we must take into account
any time-dependence of the system. In this case (in order to preserve the norm under evolution) the usual Schr\"odinger
equation must be modified to the Gong-Wong (GW) equation,~\cite{GW}
\begin{equation}
i{ \partial\psi \over \partial t} = H(t)\psi-{i\over 2}W^{-1}\dot W \psi .
\end{equation}
The additional term $\Lambda={i\over 2}W^{-1}\dot W$ (the overdot denotes
derivative in time) must be subtracted
from the Hamiltonian so that if $H$ is time-dependent, all norms as
defined with the $W$ operator, Eq.(3), be {\it independent of time.}
Note that whenever $H$ is {\it independent} of time the ${\C}$ operator
will also be. Hence, will $W$ be also independent of $t$, when the GW equation reduces to the original one of Schr\"odinger. The presence of $\Lambda(t)$ 
in the GW equation is a complicating presence, yet essential for the 
understanding of transitions in PT-symmetric quantum theory which we take 
up in the next section. Examples of the use of the GW equation will be 
presented in section IV.
\section{Selection rule for time-dependent transitions}

In this section, we will state and prove our main result. We assume that
our Hamiltonian is time-dependent in an arbitrary way, except that for
all time the system remains in the unbroken PT-phase: all of the eigenvalues
of $H(t)$, $E_n(t)$, are real for all time $t$. While still not generally
proved, evidence is strong that the eigenstates of a PT-symmetric Hamiltonian
in the unbroken phase form a complete set of states.~\cite{CB4} 
\medskip

{\bf Theorem}: If a time-dependent PT-symmetric Hamiltonian
remains in the unbroken phase, transitions between normalized energy 
eigenstates of $H$ may be made only between states of the same PT-norm. 

\medskip
In the following, the PT-norms of states will be labeled as $e_n$:
\begin{equation}
e_n=\langle\psi_n |{\P} | \psi_n \rangle = \pm 1.
\end{equation}
We start by deriving some preliminary facts. First, since ${\C}^2=1$ we have
that $0={C}\dot{\C}+\dot{\C} {\C}$ or
\begin{equation}
{\C}\dot{\C} = -\dot{\C}{\C}.
\end{equation}
Second, let $|\psi_n\rangle$ be a state of $H(t)$ (itself dependent on $t$).
Then, $e_m |\psi_m \rangle = {\C} |\psi_m\rangle$. Differentiating this
equation gives,
\begin{equation}
\dot{\C}|\psi_m\rangle = (e_m-{\C})|\dot \psi_m\rangle.
\end{equation}
Next, we note that the $\Lambda$ operator defined above can be written as
\begin{eqnarray}
\Lambda={i\over 2}W^{-1}\dot W = {i\over 2} ({\P}{\C})^{-1}{\P}\dot{\C} \cr
       =-{i\over 2}{\C}^{-1}{\P}^{-1}{\P}\dot{\C}={i\over 2}{\C}\dot{\C} \cr
       =-{i\over 2} \dot{\C}{\C}.
\end{eqnarray}
Thus, the GW equation is,
\begin{equation}
H|\Psi(t)\rangle+{i\over 2}\dot{\C}{\C}|\Psi(t)\rangle = i {\partial \Psi \over
\partial t}.
\end{equation}
We now make use of the (presumed) completeness of the eigenstates of $H$,
\begin{equation}
H(t)|\psi_n(t)\rangle = E_n(t)|\psi_n(t)\rangle,
\end{equation}
and write the solution as a superposition,
\begin{equation}
|\Psi(t)\rangle = \sum_m A_m(t) e^{-i\theta_m(t)}|\psi_m(t)\rangle,
\end{equation}
where the phase angle,
\begin{equation}
\theta_m(t)=\int_0^t E_m(t') dt',
\end{equation}
is introduced for algebraic convenience.
Substituting this ansatz into the GW equation, taking the indicated time
derivative and canceling some terms results in
\begin{equation}
-{1\over 2}\sum_m A_m e^{-i\theta_m}{\C}\dot{\C} |\psi_m\rangle =
\sum_m\dot A_m e^{-i\theta_m} |\psi_m\rangle + \sum_m A_m e^{-i\theta_m}|\dot \psi_m
\rangle.
\end{equation}
Now we take the inner product of this result with the eigenstate $|\psi_n\rangle$.
To do this we must use the correct inner product, so we first operate with 
$W={\P}{\C}$ before bringing up the bra $\langle \psi_n |$. Noting the orthogonality
of the eigenstates with respect to W: $\langle \psi_n |W|\psi_m\rangle = 
\delta_{mn}$, we obtain
\begin{equation}
e^{-i\theta_n} \dot A_n= -{1\over 2} \sum_m A_m e^{-i\theta_m} \langle \psi_n | 
{\P}(e_m-{\C}) | \dot \psi_m \rangle - \sum_m A_m e^{-i\theta_m}\langle \psi_n | 
{\P} {\C}| \dot \psi_m \rangle,
\end{equation}
where use has been made of Eq.(8). Now, two of the terms on the right hand side
can be combined as follows: The term in Eq.(15 ) $\langle \psi_n | {\P}{\C} |\dot 
\psi_m \rangle$ is an inner product of two states with $W$. However, $W$ is
self-adjoint: it can act leftward as well as rightward. Thus, acting {\it left}
produces a factor of $e_n$ - a real number - because when ${\C}$ hits an eigenstate 
of $H$ (as noted before) it simply produces the PT-norm of the respective state. 
Should this not be clear, note that 
\begin{eqnarray}
\langle \dot \psi_m | W^{\dag} |\psi_n \rangle^{*} =  
\langle \dot \psi_m | W |\psi_n \rangle^{*} 
=\langle \dot \psi_m | {\P}{\C} |\psi_n \rangle^{*}  \cr
= e_n^{*}\langle \dot \psi_m | {\P} |\psi_n \rangle^{*} \cr 
= e_n \langle \psi_n | {\P} | \dot \psi_m \rangle.
\end{eqnarray}
Collecting the derivative terms in Eq.(15) we obtain
\begin{equation}
\dot A_n = \sum_m A_m e^{i(\theta_n-\theta_m)}\big [ -{1\over 2} (e_m+e_n)
\langle \psi_n | {\P} |\dot \psi_m \rangle \big ].
\end{equation}
The above equation is our main result. Simply put, when the PT-norm $e_m$
has opposite sign to the PT-norm  $e_n$, no term with that index appears in the
sum due to the factor of $(e_m+e_n)$ and hence will not contribute to the evolution 
of the $A_n$ coefficient. In short, if the state begins, say, in a state 
of positive PT-norm it can
{\it never} evolve into any state of negative PT-norm and vice versa. This
result will be illustrated in the next section.
\section{Examples of the theorem}

For the first example, we consider the simple 2x2 matrix problem used by Bender
et al to illustrate the complications involved in PT symmetry.~\cite{CB5}
The Hamiltonian is
\begin{equation}
H=\Bigg [ \matrix { i a(t) & b \cr b & -i a(t) } \Bigg ],
\end{equation}
where $a(t)$ is a real-valued function of time and $b$ is a real constant. We choose
$a(0)=0$ as an example and suppose that the system is in an eigenstate of $H(0)$
at $t=0$, with one of the eigenvalues $E_{\pm}(0)=\pm b$. If $b^2 > a(t)^2$ for all
time, then the Hamiltonian lives in the unbroken phase for all time and the 
eigenvalues $E_{\pm}(t)$ are real at each instant of time $t\ge 0$.

For matrix models (as opposed to continuum models) $\P$ must be chosen to be a particular matrix (with the condition that the square of that matrix 
is the identity).  In this 2x2 example, we choose the ${\P}$ operator to be
\begin{equation}
{\P}=\Bigg [ \matrix {0 & 1 \cr 1 & 0 } \Bigg ].
\end{equation}
The ${\C}$ operator is then,
\begin{equation}
{\C}= {1\over \sqrt{s(t)}}\  H(t),
\end{equation}
where $s(t)=\sqrt{b^2-a(t)^2}$.
The (W-operator) normalized eigenstates of this Hamiltonian are,
\begin{equation}
|\psi_{\pm}\rangle = {1\over \sqrt{2bs(t)}} \pmatrix { b \cr \pm s(t)-ia(t)}.
\end{equation}
Once again, we substitute $|\Psi\rangle =
A_{+}|\psi_{+}\rangle + A_{-} |\psi_{-}\rangle$
into the GW equation yielding the following differential equations for the
expansion coefficients
\begin{eqnarray}
i \dot A_{+} = (s-\dot a/ 2s) A_{+} \cr
i \dot A_{-} = -(s-\dot a/2s) A_{-}
\end{eqnarray}
As the reader can see, there is no coupling between the two states, one of which
is PT-norm positive ($E_{+}$) the other which is negative.

Our second example is a 3x3 matrix Hamiltonian with two states of positive 
PT-norm and one of negative PT-norm. The Hamiltonian is explicitly
\begin{equation}
H =  \pmatrix { ia(t) & 1/\sqrt{2} & s(t) \cr
                      1/\sqrt{2} & s(t) & 1/\sqrt{2} \cr
                      s(t) & 1/\sqrt{2} & -ia(t) } ,
\end{equation}
where $s(t)=\sqrt{1+a(t)^2}$, and again we take $a(0)$ =0, when $H(0)$ has
eigenvalues $0$, $2$ and $-1$. Our $H(t)$ is PT-symmetric and in the
unbroken phase with $E_0(t)=0$, $E_{+}(t)=(s+r)/2$ and $E_{-}(t)=(s-r)/2$
where $s(t)$ is defined above and $r(t)=\sqrt{9+a(t)^2}$. We choose the parity operator to be
\begin{equation}
{\P}= \pmatrix {0 & 0 & 1 \cr 0 & 1 & 0 \cr 1 & 0 & 0} .
\end{equation}
Given the Hamiltonian above, one can compute its
(un-normalized) eigenstates, and given the $\P$ operator one can compute the sign of the PT norm of each of those
eigenstates.  The results are as follows:
\begin{eqnarray}
|\phi_0\rangle = \pmatrix { s+ia \cr -\sqrt{2} \cr s-ia} \quad \hbox{ positive PT-norm} \cr
|\phi_{+}\rangle =\pmatrix {E_{+}(t) + s +ia(t) \cr \sqrt{2}(E_{+}^2 -1) \cr E_+ s-ia(t)}
\quad \hbox {positive PT-norm} \cr
|\phi_{-}\rangle = \pmatrix {E_{-}(t) + s +ia(t) \cr \sqrt{2}(E_{-}^2-1) \cr E_{-} +s-ia(t)}
\quad \hbox {negative PT-norm}
\end{eqnarray} 
Here $|{\phi_0}>$ (respectively $|{\phi_+}>, \, |{\phi_-}>$) is the eigenstate with eigenvalue $E_0$ (respectively
${E_+} , \, {E_-}$).
The ${\C}$ operator must map $|{\phi_0}>$ into $|{\phi_0}>$ and $|{\phi_+}>$ into $|{\phi_+}>$ and 
$|{\phi_-}>$ into $-|{\phi_-}>$.  It then follows that $\C$ is given by the expression
\begin{equation}
{\C}= I + {(s+r)^2 \over 4r} H - {s+r\over 2r} H^2.
\end{equation}
That the ${\C}$ operator is a simple polynomial function of $H$ in these 
examples is due to the finite sizes of the matrix Hamiltonians.

We suppose that the state is the $E_0=0$ state at $t=0$ and ask what is
the probability that the state is any one of the three at an arbitrary
time in the future? To answer this question, the above eigenstates must
be correctly normalized by $W(t)$. Let $|\chi_n\rangle$ denote 
the nth state above $W(t)$-normalized at $t=0$. Then the probability that 
the nth state obtains at a later time is
\begin{equation}
P_n(t)=|\langle \chi_n | W(t) |\Psi_0(t)\rangle|^2
\end{equation}
where $n=0,\pm$ and $|\Psi_0 \rangle$ is the solution of the GW equation
with initial conditions appropriate for the $E_0=0$ starting state.   
To illustrate, we take two choices of $a(t)$ 
\begin{eqnarray}
a_1(t)=\alpha(1-e^{-\beta t}) \cr
a_2(t)=\alpha \sin{\omega t}.
\end{eqnarray}

\begin{figure}
\centering
\includegraphics[width=4.5in,angle=-90]{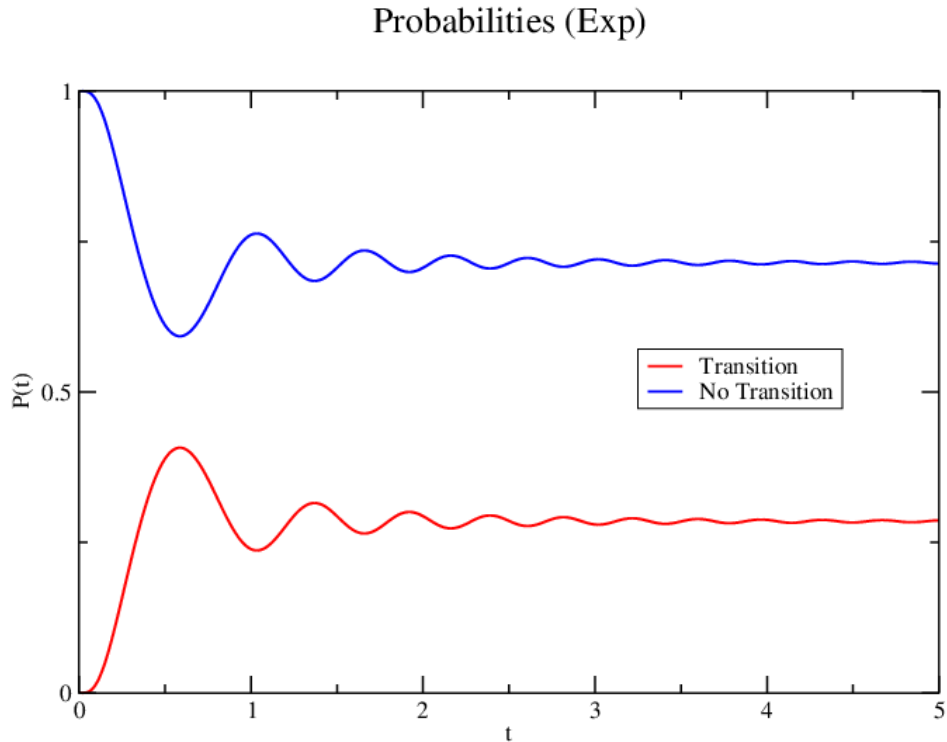}
\caption{Probabilities of transition or no transition for the function
$a_1(t)$ in Eq.(28) with $\alpha =20$ and $\beta =1/2$.}
\label{Fig.1}
\end{figure}
After solving the GW equations (5) numerically (double
precision) we calculate the above probabilities. Indeed, the probability
of finding the system in the $E_{-}$ state with negative PT-norm is
zero to machine precision for all $t$. The probabilities of revealing the
state to be $E_0$ (no transition) or the $E_{+}$ (transition occurring)
are shown in the figures. The results are shown in Figure 1 (for ${a_1}(t) $ with 
$\alpha = 20$ and $\beta =1/2$) and Figure 2 (for ${a_2}(t)$ with $\alpha =2.5 $ and 
$\omega = 2.5$).  The parameters, of course, can be varied but our numbers were picked 
to produce behaviors typical of the system.  Figure 1 shows the transition/no transition
probabilities for the exponential choice above; there are mild oscillations
as they approach a limit. In figure 2, we show the corresponding
results for the $a_2(t)$ choice. In this case, the probabilities show a
complex beat structure.

\begin{figure}
\centering
\includegraphics[width=4.5in,angle=-90]{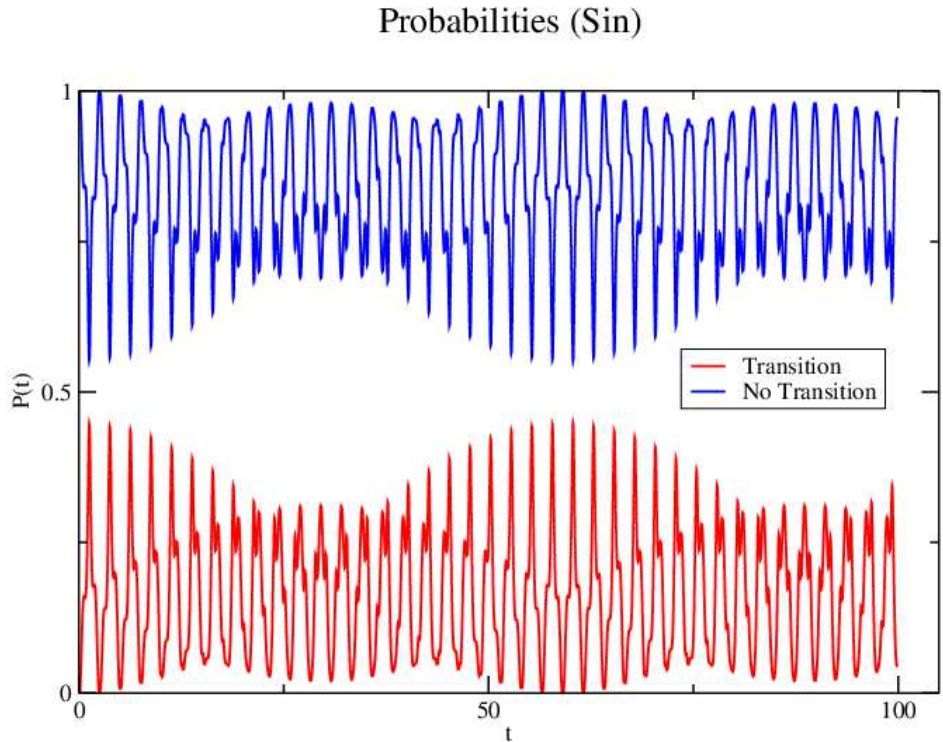}
\caption{Probabilities of transition or no transition for the function
$a_2(t)$ in Eq.(28) with $\alpha =2.5$ and $\omega = 2.5$.}
\label{Fig.2}
\end{figure}

Our last example is a simple solvable continuum model. Consider the
Hermitian Hamiltonian
\begin{equation}
H=1/2 \, p^2 + 1/2\, x^2 + g(t) x
\end{equation}
which is a shifted harmonic oscillator with a time-dependent shift. We
take $g(0)=0$ so that the linear perturbation turns on at $t=0$. 
The matrix elements between oscillator states $\langle \phi_n|x|\phi_m\rangle$ 
do not vanish when $m=n\pm1$ and thus transitions between states are induced by
this time-dependent perturbation as long as the state indices differ
by at most one unit. Will this phenomenon persist if the perturbation
is PT-symmetric? Consider
\begin{equation}
H=1/2 \, p^2 + 1/2\, x^2 + i g(t) x,
\end{equation}
where again $g(0)=0$. This PT-symmetric Hamiltonian is exactly solvable.
The energy eigenvalues are $E_n=n+1/2+g(t)^2/2$, $n=0,1,2\dots$ with
corresponding eigenstates
\begin{equation}
\psi_n=\phi_n(x+ig)=e^{-g(t)\hat p} \phi_n(x),
\end{equation}
where again the $\phi_n(x)$ are the eigenstates of the harmonic oscillator,
and $\hat p$ is the momentum operator. The ${\C}$ operator for this system
is
\begin{equation}
{\C}={\P} e^{2g\hat p}=e^{-2g\hat p}{\P}.
\end{equation}
Will there be transitions between oscillator states (defined at $t=0$)
induced by the perturbation turning on?  We already have a selection
rule for these transitions which is our Eq.(17). One factor on the right
hand side sum is $(e_n+e_m)/2$. In order for the term labeled by $m$ to
appear in the sum, it must be that $e_n$ and $e_m$ have the same sign, either
both plus or both minus. Thus, if $n=0$ say (the ground state of the
oscillator) then $m$ can only be $0,2,4,\dots$, states with the same
PT-norm. However, there is another factor $\langle \psi_n | {\P} | \dot
\psi_m \rangle$ inside the sum. For the simple system above, we may
calculate this factor:
\begin{eqnarray}
\langle \psi_n | {\P} | \dot\psi_m \rangle =\langle \phi_n| e^{-g\hat p} {\P}
{d\over dt} e^{-g \hat p} |\phi_m\rangle \cr
=-\dot g(t) \langle \phi_n| e^{-g\hat p} e^{+g\hat p} {\P}\hat p | \phi_m 
\rangle \cr
=+\dot g(t) (-1)^m \langle \phi_n | \hat p | \phi_m \rangle .
\end{eqnarray}
Hence, immediately in Eq.(17) we have
\begin{equation}
\dot A_n=0.
\end{equation}  

The last result follows because the momentum operator, $\hat p$, can connect
only those states whose quantum numbers differ by at most unity, which
violates the selection rule. Thus, for this system, once it is in an
eigenstate of energy it {\it cannot transition to a different eigenstate
of energy}. This effect makes the system entirely trivial and much
different from its Hermitian cousin.

One may ask if this non-transition effect persists for more interesting
systems, such as the harmonic oscillator with an $ig(t) x^3$ perturbation, a
system which has been studied but for which there are no analytic results
available; the ${\C}$ operator has been calculated in perturbation 
theory.~\cite{CB3}
Also, what transitions are allowed for the pure $igx^3$ oscillator
for which it has been proved that there are all real eigenvalues.~\cite{IX} 
Does the selection rule persist in PT-symmetric quantum field theory? If so
what restrictions on physical phenomena does it inhibit or allow?
\newpage
{\bf Acknowledgements}
The work of DG is supported in part by NSF grant PHY-1505565  to Oakland University.

\end{document}